\documentclass[%
 reprint,
 amsmath,amssymb,
 aps,
]{revtex4-1}

\usepackage{graphicx}
\usepackage{dcolumn}
\usepackage{bm}
\usepackage{color}

\newcommand{\ie}{\textit{i}.\textit{e}.}

\begin{document}

\title{Nuclear fission reaction simulations in compact stars}

\author{Alex Deibel}\email{deibelalex@gmail.com}
\affiliation{Department of Astronomy, Indiana University,
                  Bloomington, IN 47405, USA}

\author{M. E. Caplan}
 \email{mecapl1@ilstu.edu}
\affiliation{
 Department of Physics, Illinois State University, Normal, IL 61790, USA 
}%

\author{C. J. Horowitz}\email{horowit@indiana.edu}
\affiliation{Center for Exploration of Energy and Matter and
                  Department of Physics, Indiana University,
                  Bloomington, IN 47405, USA}

\date{\today}

\begin{abstract}
Type\textrm{--}Ia supernovae are powerful stellar explosions that provide important distance indicators in cosmology.  Recently, we proposed a new SN Ia mechanism that involves a nuclear fission chain-reaction in an isolated white dwarf [PRL {\bf 126}, 1311010].  Here we perform novel reaction network simulations of the actinide-rich first solids in a cooling white dwarf. The network includes neutron-capture and fission reactions on a range of U and Th isotopes with various possible values for $^{235}$U enrichment.  We find, for modest $^{235}$U enrichments, neutron-capture on $^{238}$U and $^{232}$Th can breed additional fissile nuclei so that a significant fraction of all U and Th nuclei may fission during the chain-reaction. Finally, we compute the energy release from the fission chain-reaction for various uranium enrichments; a novel result that is a necessary input for thermal diffusion simulations of carbon ignition.


\end{abstract}

\maketitle

\section{Introduction}

Type\textrm{--}Ia supernovae (SN Ia) are widely used distance indicators in cosmology \cite{Abbott_2019,SN_cosmology,Sullivan2010}, but significant tension remains between the Hubble constant determined from SN Ia and the value determined from other data \cite{Riess:2016,Riess_2021,di_valentino_2021}. Despite the importance of SN Ia for cosmology, their progenitor systems and explosion mechanism are still somewhat uncertain.

Traditionally, SN Ia are thought to involve the thermonuclear explosion of a C/O white dwarf (WD) in a binary system. Here the companion is either a conventional star (single-degenerate mechanism) or another WD (double-degenerate mechanism) \cite{2012NewAR..56..122W,hillebrandt2013understanding,RUIZLAPUENTE201415}. Recently we proposed an additional SN Ia mechanism that may occur in {\it isolated} WDs \cite{PhysRevLett.126.131101,fission2} wherein the cooling WD core rapidly precipitates a fission-critical uranium crystal within $\lesssim 30\, \mathrm{s}$. If a fission chain-reaction proceeds in the crystal, it is unknown, a priori, the fraction of uranium consumed or the resulting energy release.

In the present paper we perform nuclear reaction network simulations of fission chain-reactions in compact stars.   Our goal is to determine the fraction of fissile fuel consumed during a fission chain-reaction and the resulting energy release and how this depends on the uranium enrichment $f_5$ \textrm{--} the fraction of all uranium that is $^{235}$U. This is important input for thermal diffusion simulations of carbon ignition in an isolated WD.  We present these simulations in a separate paper \cite{bombcode} where we find that carbon ignition is likely, at high densities.

Our simulations are novel, apparently the first such calculations for a compact star.  
To provide context, we briefly review fission chain-reactions in conventional nuclear reactors and nuclear weapons in Sec. \ref{sec.contect}.  Our reaction network formalism is described in Sec. \ref{sec.formalism}, results presented in Sec. \ref{sec.results} and we conclude in Sec. \ref{sec.conclusions}.


\section{Context\label{sec.contect}}

 A fission reaction in a compact star is a unique hybrid between a nuclear reactor and a nuclear weapon.  Like a nuclear weapon the chain-reaction is expected to proceed extremely rapidly.  A WD is degenerate, however, and the temperature can rise without a large increase in pressure.  As a result, the system does not rapidly disassemble as in a nuclear weapon.  This allows time \textrm{--} as in a nuclear reactor \textrm{--} for fertile isotopes such as $^{238}$U or $^{232}$Th to capture neutrons and breed additional fissile material.

 Many conventional nuclear reactors slow neutrons to (terrestrial) thermal energies to take advantage of the large fission cross section of $^{235}$U at low energies. In the WD core, however, the temperature is of order $\sim 1\, \mathrm{keV}$ and the $^{235}$U fission cross section is much smaller.  Even if plenty of light nuclei are present to moderate the neutrons, the neutron energy will only be reduced to the $\sim 1\, \mathrm{keV}$ ambient temperature. Therefore, unlike a terrestrial nuclear reactor, a stellar system can not take advantage of the large low-energy $^{235}$U cross section.  
 
Terrestrial nuclear weapons, on the other hand, disassemble extremely rapidly because of the large energy release.  This necessitates using fast neutrons in the chain-reaction because slow neutrons simply take too long to cause additional fissions.  By the time a slow neutron arrives to cause another fission the fuel may have been blown apart.  
This reliance on fast neutrons requires the use of highly enriched uranium or plutonium in a nuclear weapon. 
 
The necessary uranium enrichment may be reduced if it is possible for the chain-reaction to breed additional fissile nuclei.  For example $^{238}$U in a nuclear reactor can capture a neutron to become $^{239}$U that in turn beta decays twice to produce $^{239}$Pu.  
Therefore, a fission chain-reaction in a star could breed some of its nuclear fuel as the reaction progresses.

  We now begin our study by discussing the composition of the first solids to form as a WD cools.  Next, we list the fission and neutron-capture reactions that are included in our network simulations and present the results for composition and energy release as a function of time.  We end with a discussion of sensitivity to uranium enrichment and to the $^{238}$U fission cross section.  We conclude that for modest enrichments a large fraction of the available fuel is expected to fission producing a large energy release that could ignite carbon burning.



\section{Formalism}
\label{sec.formalism}

{\it Initial abundances:} The composition of the first solids to form as material in a WD just starts to crystallize has been studied using free energy models and with molecular dynamics simulations \cite{PhysRevLett.126.131101,fission2}.  The material is U and Th rich since these elements have the highest charge $Z$.  In addition some Pb is present because the solar system abundance of Pb is 100 times that of U.  The initial abundance in nuclei per baryon $Y_i=n_i/n_b$ are listed in Table~\ref{Table1}.  Here $n_i$ is the number density of species $i$ and $n_b$ is the baryon density.

In addition to heavy nuclei, some C and O may be present in the first solids.  Elastic scattering from the light C and O nuclei can lower the energy of fission neutrons.  At this time the amount of C and O is uncertain and may be zero.  For simplicity in this first study, we assume there is no C and O present.  As a result there will be little moderation of the initial neutron energies.  The fission spectrum has a most probable energy near $\approx 1\, \mathrm{MeV}$.  In this paper we simply assume all neutrons have an energy of $1\, \mathrm{MeV}$ and evaluate all cross sections at this energy.  This assumption of mono-energetic neutrons greatly simplifies thermally averaged reaction rates that are proportional to the cross section times the relative velocity,
\begin{equation}
    \langle \sigma(E)v\rangle \approx \sigma({\rm 1\, MeV})v_0\, ,
\end{equation}
with $v_0$ the velocity of a $1\, \mathrm{MeV}$ neutron.  Furthermore, this thermal average is independent of temperature.

\begin{table}[t]
\caption{\label{Table1} Initial abundances $Y_i$, electron fraction $Y_e$, and baryon number density $n_b$.}
\begin{tabular*}{0.45\textwidth}{c c c c c} \hline \hline
U & Th & Pb & $Y_e$ & $n_b$ ($\mathrm{cm}^{-3}$)\\ \hline
$1.3\times 10^{-3}$& $9.51\times 10^{-4}$ & $2.28\times 10^{-3}$& 0.391 & $6.02 \times 10^{31}$ \\  
\hline \hline
\end{tabular*}
\end{table}

\begin{table}[t]
\caption{\label{Table2} Cross sections for neutron absorption $(n,\gamma)$ and fission reactions for 1 MeV neutrons on Th and U isotopes.  Also listed is the average number of neutrons emitted per fission $\bar\nu$.  Data from \cite{NNDC}. }
\begin{tabular*}{0.2345\textwidth}{c c c c  } \hline \hline
Isotope & $\sigma_{n,\gamma}$ (b) & $\sigma_f$ (b)&$\bar\nu$ \\ \hline
$^{232}$Th & 0.14 & 0.0013 & 2.18 \\
$^{233}$Th & 0.068 & 0.094 &  2.69 \\
$^{235}$U & 0.11 & 1.20 & 2.53 \\
$^{236}$U & 0.17 & 0.36 & 2.49 \\
$^{237}$U & 0.082 & 0.68 & 2.57 \\
$^{238}$U & 0.13 & 0.014 & 2.69 \\
$^{239}$U & 0.097 & 0.38 & 2.91 \\
$^{240}$U & 0.086 & 0.007 & 2.69 \\
$^{241}$U & 0.17 & 0.24 & 2.88 \\
 \hline \hline
\end{tabular*}
\end{table}

We consider neutron-capture $(n,\gamma$) and neutron-induced fission reactions on the U and Th isotopes.  We use 1 MeV cross sections from the ENDF 2011 data set available at the National Nuclear Data Center \cite{NNDC}, see Table \ref{Table2}.  Our reaction network has $^{232}$Th, $^{233}$Th, and $^{234}$Th isotopes and U isotopes from $^{235}$U to $^{242}$U.  In addition we have neutrons and fission fragments.  We do not distinguish different possible fission fragments and simply assume that any fission will produce two fragments.  Our network has a total of 13 species consisting of 3 Th isotopes, 8 U isotopes, $n$ and fission fragments.

There are simple equations for the change in abundance $dY_i/dt$ from a given reaction \cite{Lippuner_2017}.  For example the change in neutron abundance $Y_n$ from the fission of nucleus $A$ of abundance $Y_A$ is
\begin{equation}
\frac{dY_n}{dt}=(\bar\nu-1)n_b\sigma_fv_0 Y_n Y_A\, . 
\label{eq.dydt}
\end{equation}
Here each fission produces an average number of neutrons $\bar\nu$ (see Table \ref{Table2}, and one neutron was absorbed to cause the fission).  The fission also increases the abundance of fission fragments $Y_{\rm ff}$ where we include a factor of two for the two fragments,
\begin{equation}
\frac{dY_{\rm ff}}{dt}=2 n_b\sigma_fv_0 Y_n Y_A\, . 
\label{eq.dyffdt}
\end{equation}
Likewise neutron absorption on nucleus $A$ decreases its abundance and increases the abundance of nucleus $A+1$,
\begin{equation}
    \frac{dY_{A+1}}{dt}=-\frac{dY_A}{dt}=n_b\sigma_{n,\gamma}v_0 Y_n Y_A\, .
    \label{eq.dyAdt}
\end{equation}
We sum terms with these forms over all of the $(n,\gamma)$ and fission reactions in Table~\ref{Table2}.

\begin{table}[tb]
\caption{\label{Table3} Four cases of enrichment $f_5$, fission cross section of $^{238}$U, and neutron absorption cross section of fission fragments $\sigma_{\rm ff}$.  Also listed are the initial $n$ abundance $Y_n$, the fission heating $S$, the fraction of U and Th that fission, and the final temperature $T_f$.}
\begin{tabular*}{0.4476\textwidth}{c c c c c c c c c } \hline \hline
Case & $f_5$ & $\sigma_f(^{238}$U)& $\sigma_{\rm ff}$& $Y_n$ & S & U & Th & $T_f$ \\ 
 & & (b) & (b) & & (MeV/A) & \% & \% & $10^{9}$ K \\ \hline
 A & 0.14 & 0.014 & 0 &$10^{-9}$ & 0.026& 9.6 & 0.4& 1.6\\
 B & 0.14 & 0.04 & 0 & $10^{-6}$ & 0.358 & 95 & 58& 5.9 \\
 C & 0.20 & 0.014 & 0 & $10^{-6}$& 0.356 & 94 & 58& 5.9 \\
 D & 0.14 & 0.04 & 0.01 & $10^{-6}$ & 0.325 & 89 & 49& 5.6 \\
\hline \hline
\end{tabular*}
\end{table}

The initial abundance of $^{235}$U is $f_5Y_u$ with $Y_u$ the uranium abundance from Table \ref{Table1} and $f_5$ the uranium enrichment.  Likewise the $^{238}$U abundance is $Y_u(1-f_5)$.  Finally for the initial neutron abundance one can use any small seed value.  The cases explored are listed in Table \ref{Table3}.

The energy released per fission is about $200 \, \mathrm{MeV}$, or $100\, \mathrm{MeV}$ per fission fragment.  Therefore the heating rate in MeV per baryon per time is 
\begin{equation}
    \dot S= (100\, {\rm MeV})\, dY_{\rm ff}/dt,
\end{equation}
and the total energy released by a time $t_{\rm final}$ in MeV per baryon is 
\begin{equation}
 S=(100\ {\rm MeV})\, Y_{\rm ff}(t_{\rm final})  \ .
\end{equation} 
The large fission energy release will raise the temperature of the system.  We assume the reaction proceeds at constant pressure.   Previously we calculated the heat capacity at constant pressure and obtained the final temperature $T_f$ \cite{fission2},
\begin{equation}
 T_f\approx \Bigl(\frac{8\epsilon_F S}{5\pi^2Y_e}\Bigr)^{1/2}\, .
\end{equation}
Here $\epsilon_F$ is the electron Fermi energy and $Y_e$ is the electron fraction, see Table~\ref{Table1}.

\begin{figure*}[bt!]
\centering  
\includegraphics[width=0.48\textwidth]{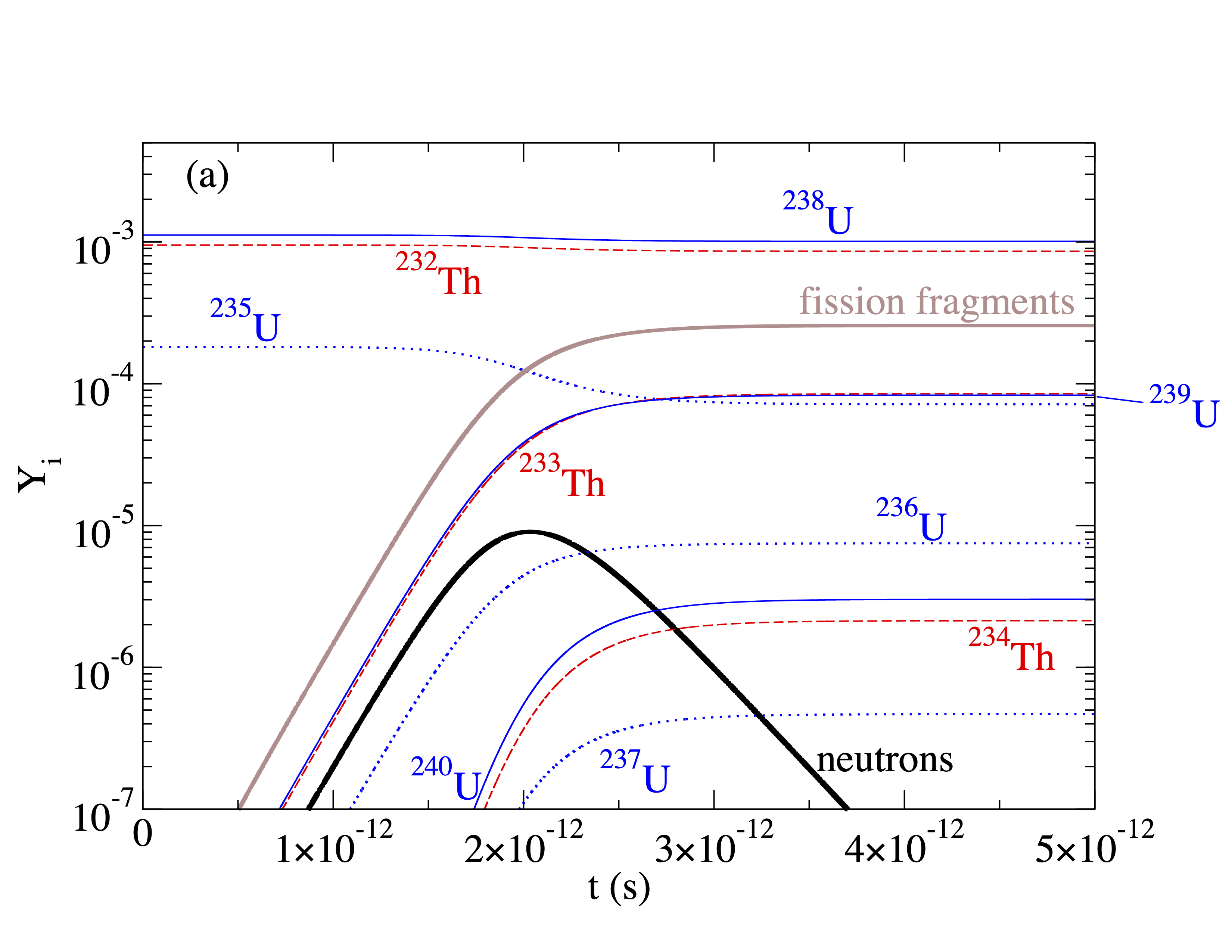}
\includegraphics[width=0.48\textwidth]{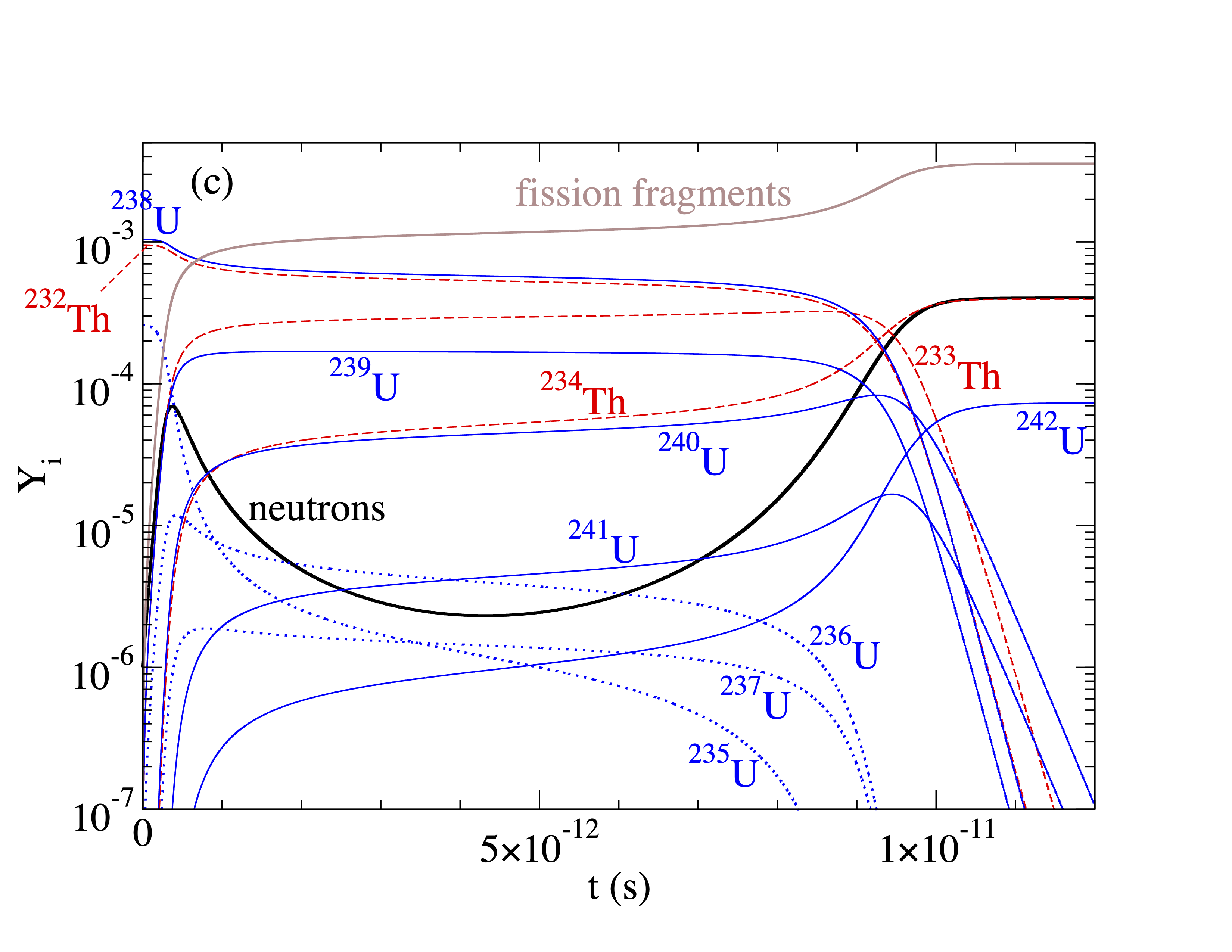}
\includegraphics[width=0.48\textwidth]{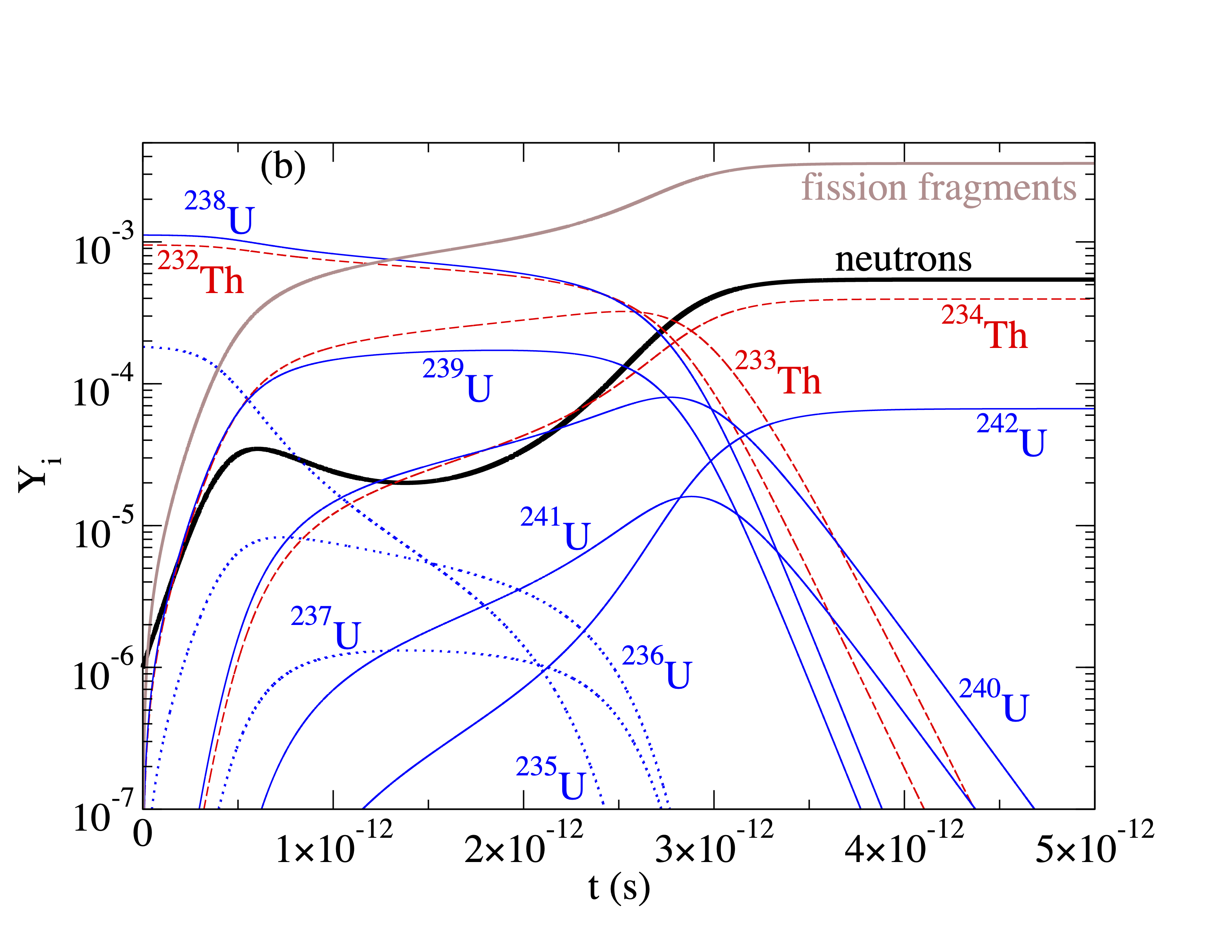}
\includegraphics[width=0.48\textwidth]{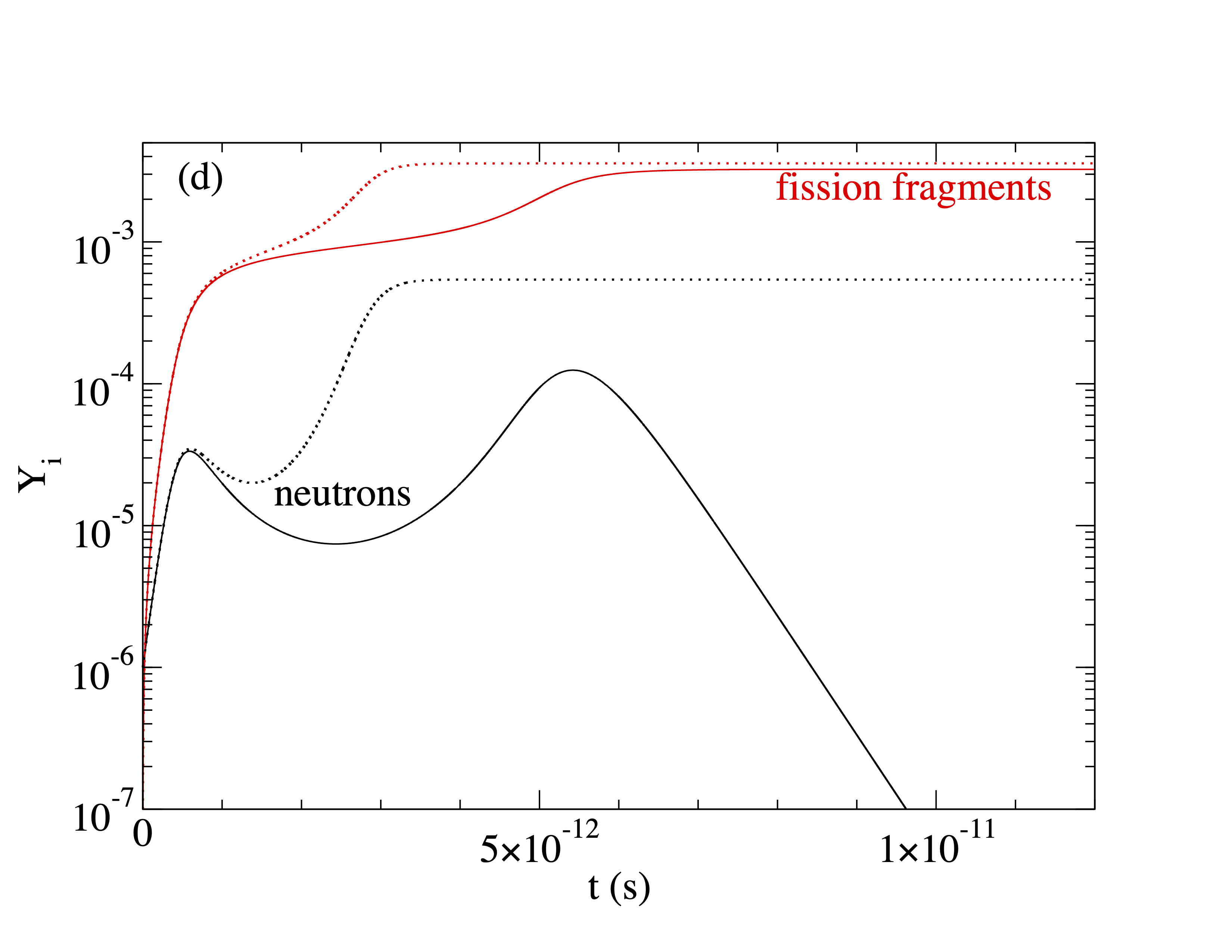}
\caption{\label{Fig1} Abundance per baryon $Y_i$ versus time for Case A in panel (a), B in (b) and C in (c).  Panel (d) shows abundance per baryon of neutrons (black curves) and fission fragments (red curves) vs time.  Results are shown for Case B without neutron capture on fission fragments (dotted) and for Case D with capture (solid), see Table~\ref{Table3}.}	
\end{figure*} 

\section{Results\label{sec.results}}


\subsection{Abundance evolution}

We now present results for three cases of enrichment and $^{238}$U fission cross section as listed in Table ~\ref{Table3}. In all cases the chain-reaction proceeds rapidly, in less than $10^{-11} \, \mathrm{s}$, as shown in Fig.~\ref{Fig1}.  This is because of the high density of the system, the large neutron cross sections, and the high velocity of $1\, \mathrm{MeV}$ neutrons.
In general, the reaction proceeds in two stages.  In the first stage neutrons from $^{235}$U fission transform or breed some $^{238}$U and $^{232}$Th nuclei into more easily fissionable $^{239}$U and $^{233}$Th.  In the second stage, or breeder reaction, most of the original U and a significant fraction of the Th fission. 

In Case A we use $f_5=0.14$ and the unmodified $1\, \mathrm{MeV}$ fission cross section for $^{238}$U of $0.014\, \mathrm{b}$.  One needs an enrichment of at least $f_5\approx 0.12$ for the system to be critical.  If $f_5$ is smaller than that no chain-reaction will take place.  If $f_5$ is only slightly larger than 0.12 we expect the chain-reaction to burn a small amount of $^{235}$U until the system becomes sub-critical and the chain reaction stops.  This will only release a small amount of fission heating.


In Fig.~\ref{Fig1} (a) we show results for $Y_i$ versus time for Case A.  The system fissions $\approx 9.6, \%$ of the total U.  This includes over half of the original $^{235}$U and only a small amount of $^{238}$U.  Only a small amount of Th fissions near $\approx 0.4 \, \%$.  We see that the neutron abundance rises exponentially with time until enough $^{235}$U has been burned so that the system is no longer critical. After that $Y_n$ decreases as the remaining neutrons are captured.

\begin{figure}[tb]
\centering  
\includegraphics[width=0.48\textwidth]{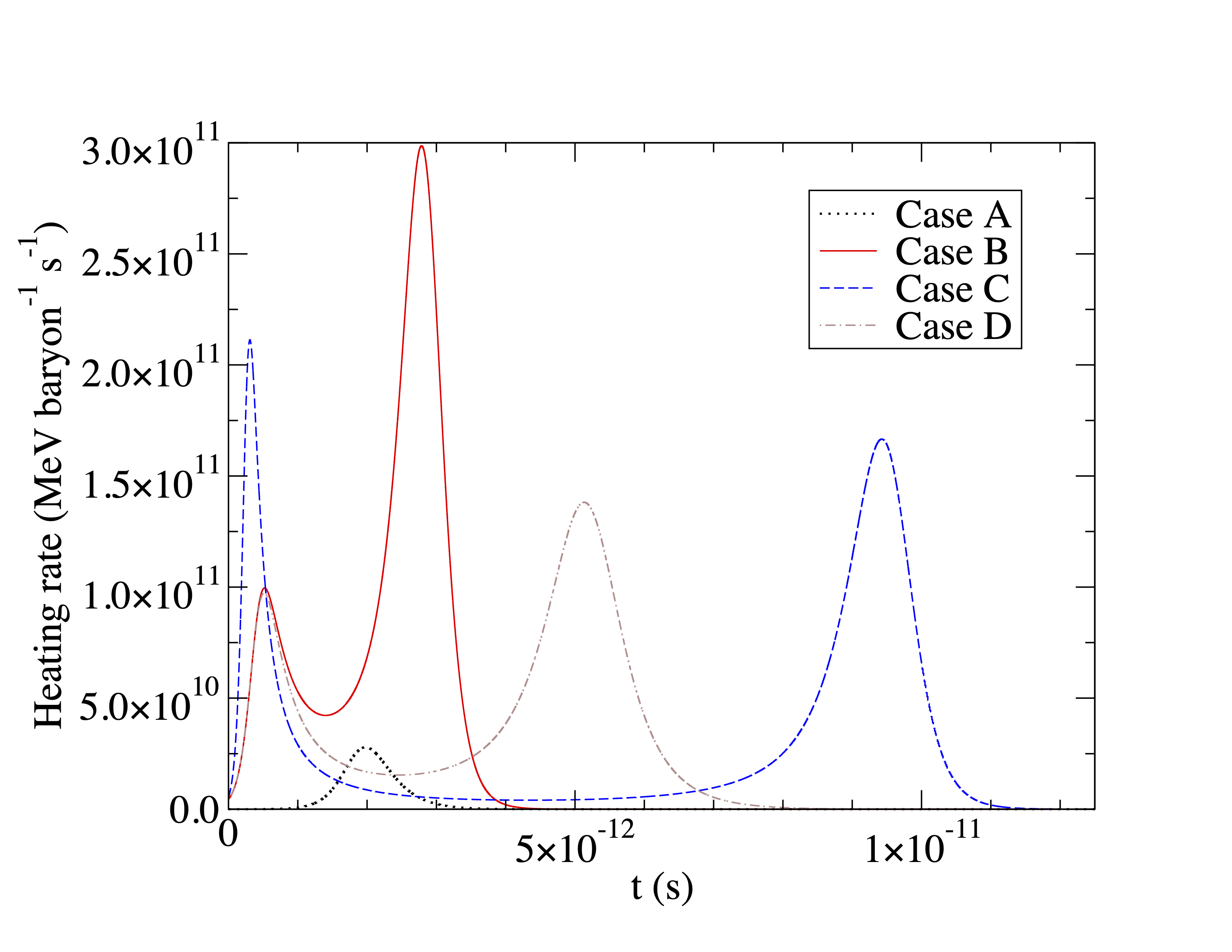}
\caption{\label{Fig2} Heating rate per baryon $\dot S(t)$ versus time for the four cases in Table \ref{Table3}. }	
\end{figure}

\begin{figure}[tb]
\centering  
\includegraphics[width=0.48\textwidth]{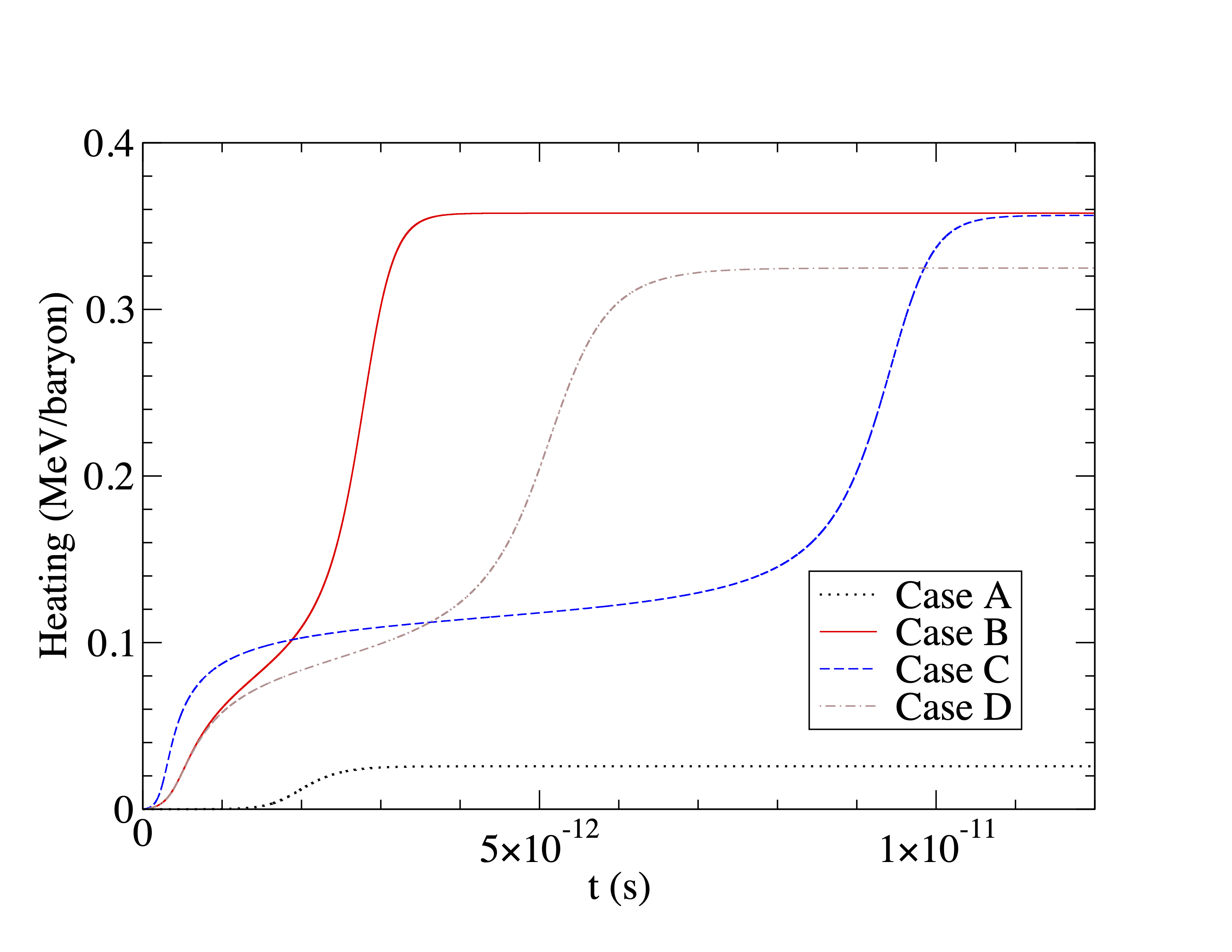}
\caption{\label{Fig3} Total fission heating per baryon $S(t)$ versus time for the four cases in Table~\ref{Table3}. }	
\end{figure}



Results are very sensitive to the small $^{238}$U fission cross section.  This is $0.014\, \mathrm{b}$ at $1\, \mathrm{MeV}$ but rises rapidly at higher energies.  A detailed calculation averaging the energy-dependent cross section over the neutron fission spectrum may give a larger value.  Alternatively, as the temperature of the medium rises, $^{238}$U nuclei will occupy a range of excited states and these may have higher fission cross sections for $1\, \mathrm{MeV}$ neutrons.  For example, Zhu and Pei calculate that the spontaneous fission half life of $^{240}$Pu decreases by 12 orders of magnitude as the temperature is increased from $0\,\mathrm{MeV}$ to $0.1\, \mathrm{MeV}$  \cite{PhysRevC.94.024329}. In Case B we use $\sigma_f=0.04\, \mathrm{b}$ instead of $0.014\, \mathrm{b}$ for $\sigma_f(^{238}$U).  The results in Fig.~\ref{Fig1} (b) show dramatic differences from Case A.  The reaction now proceeds in two stages.  First mostly $^{235}$U fissions.  This releases enough neutrons so the $n$ capture converts both $^{232}$Th and $^{238}$U into odd A nuclei with significant fission cross sections.  In the second or breeder reaction stage these nuclei fission.  As a result fully 95\% of the U and 58\% of the Th fissions.  Note that in a terrestrial nuclear reactor there is time for $^{239}$U to decay to $^{239}$Pu.  Here there is not enough time and $^{239}$U instead can be used directly as a fuel.

Alternatively, even if the $^{238}$U cross section is only $0.014\, \mathrm{b}$, one can obtain a breeder reaction stage by modestly increasing $f_5$.  Case C has $\sigma_f=0.014$ b but uses an enrichment of $f_5=0.20$ instead of 0.14.  The results in Fig.~\ref{Fig1} panel (c) show two well-separated reaction stages and the fission of a large fraction of available nuclei similar to Case B.
 
We have assumed that the fission fragments are essentially inert.  While there may not be time for beta-decay, the fission fragments could capture neutrons.  Many fission fragments have $(n,\gamma)$ cross sections for $1\, \mathrm{MeV}$ neutrons of order $\sim 0.01 \,\mathrm{b}$ \cite{NNDC}.  Therefore to explore this we simply assign all fission fragments a $0.01\, \mathrm{b}$ capture cross section for Case D, see Table~\ref{Table3}.  This case is otherwise identical to Case B.  Note that we are not keeping track of the identity of each fission fragment so when a fragment captures a neutron its identity does not change.  Therefore, neutron-capture on fission fragments simply acts as a neutron sink.

Figure \ref{Fig1} panel (d) shows that capture on fission fragments somewhat reduces the abundance of neutrons.  Indeed Case B has a finite abundance of neutrons remaining.  This artificial result reflects the limitations of the reaction network.  In Case D all neutrons are eventually captured.  The reduction in neutrons slows the production of fission fragments somewhat.  However by the time the reaction is over, the total number of fission fragments and therefore the total fission energy released is only slightly smaller in Case D with capture than originally in Case B.  
 
\begin{figure*}[ht!]
\centering  
\includegraphics[trim=0 105 310 45,clip,height=5cm,keepaspectratio]{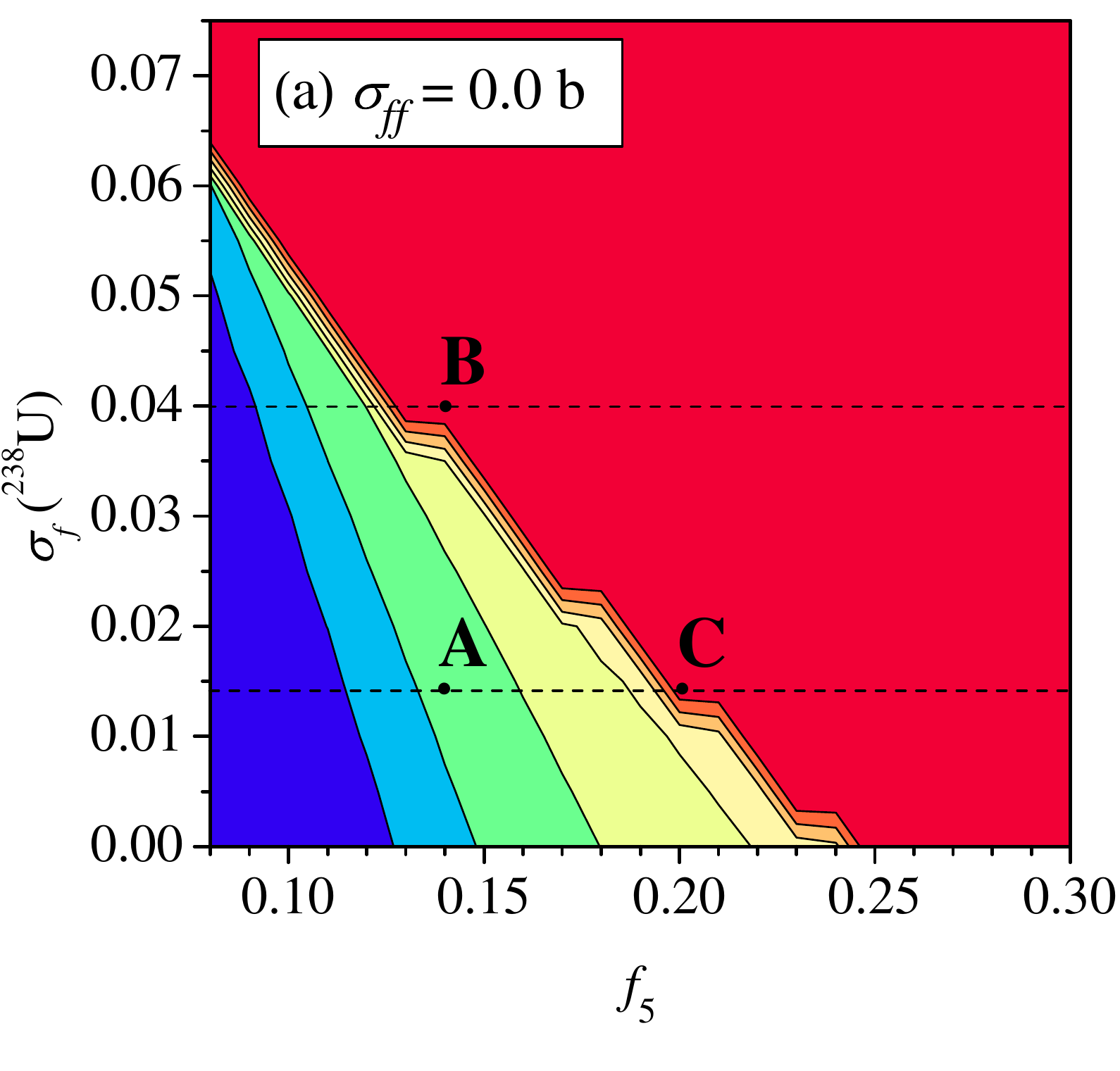} 
\includegraphics[trim=0 105 310 45,clip,height=5cm,keepaspectratio]{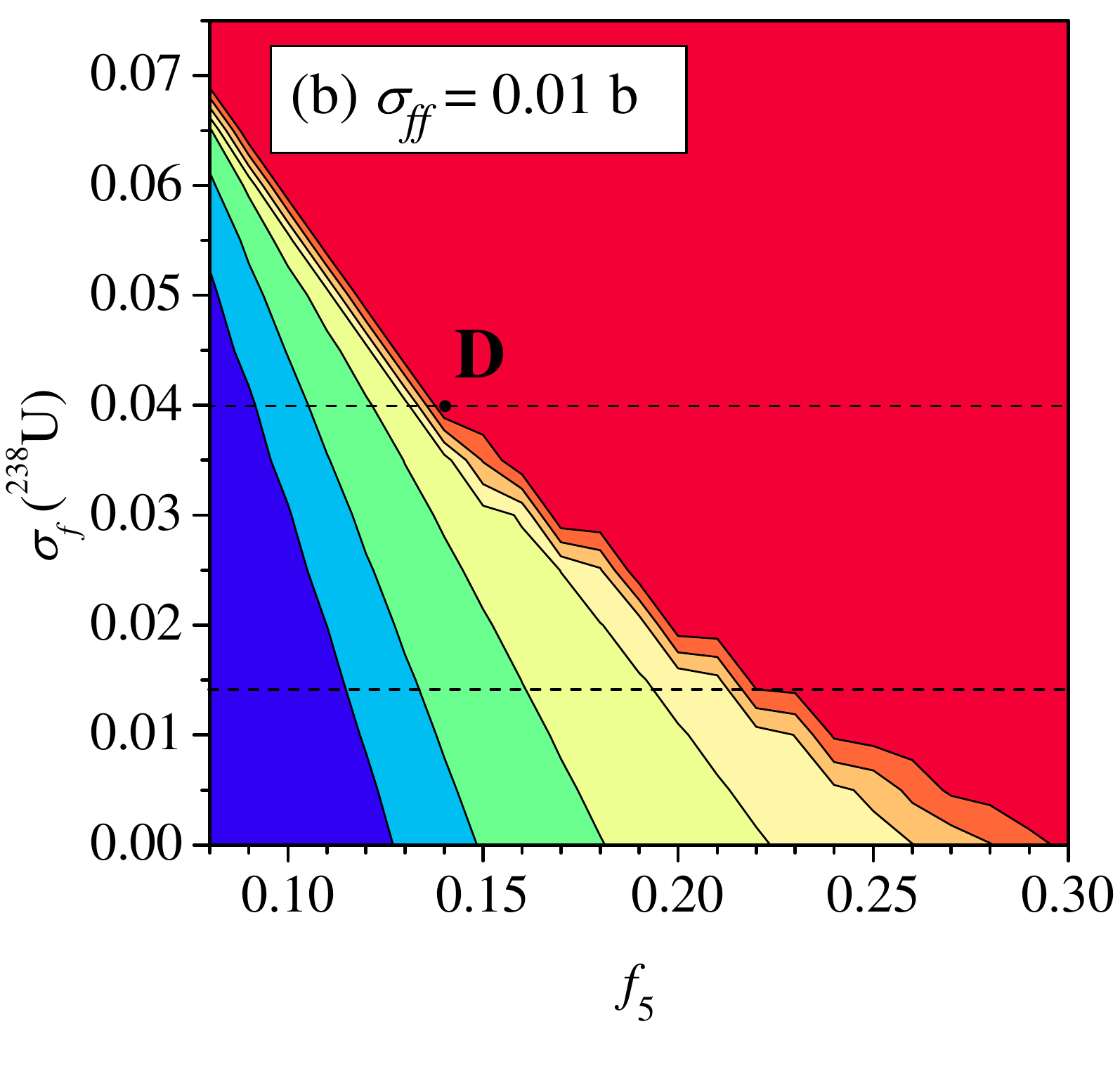}
\includegraphics[trim=0 105 213 45,clip,height=5cm,keepaspectratio]{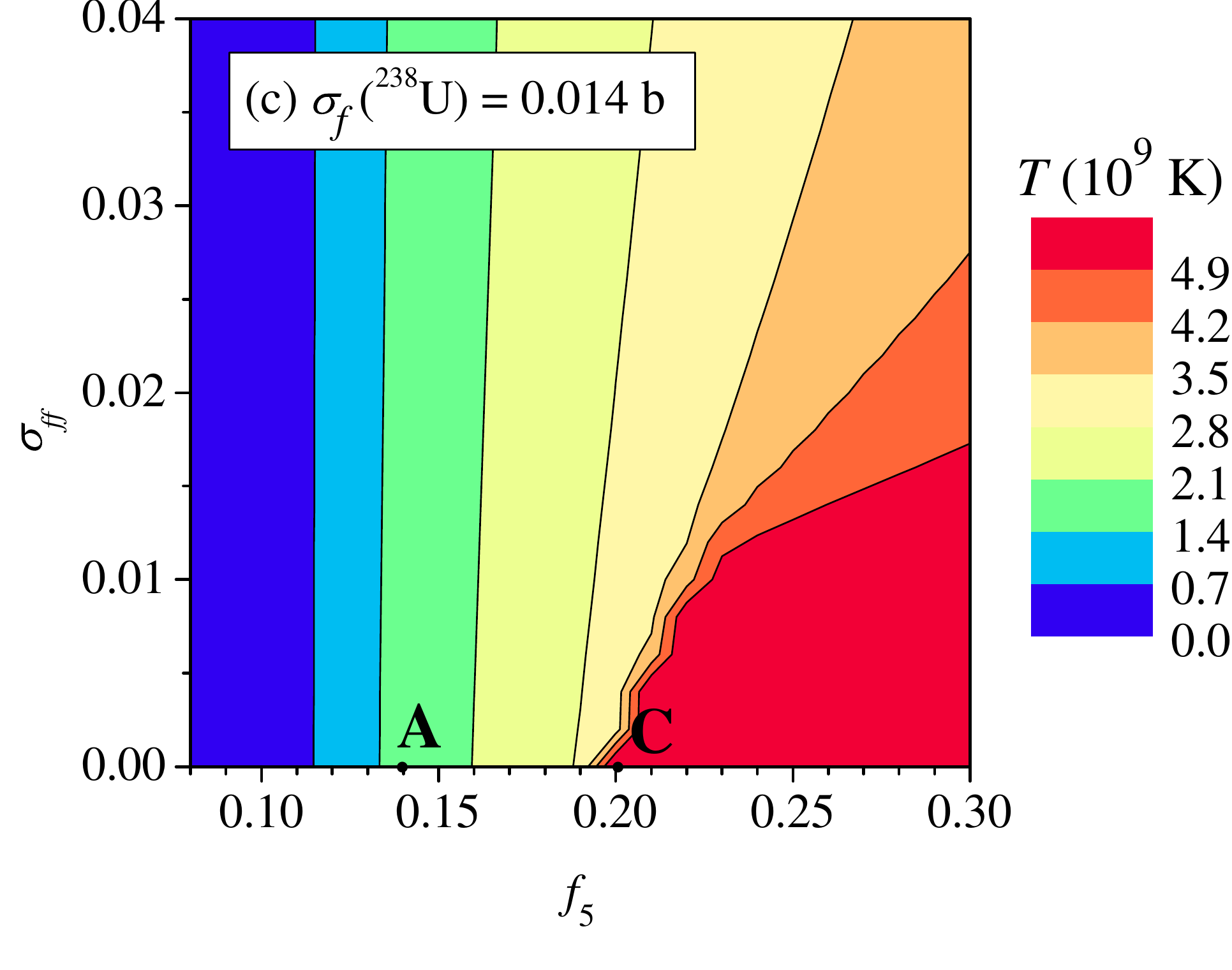}
\caption{\label{Fig6} Final temperature as a function of $f_5$ and $\sigma_f(^{238}$U) using constant $\sigma_{\rm ff}$ = 0 (left) and $0.01\, \mathrm{b}$ (center). Right panel shows the final temperature as a function of $f_5$ and $\sigma_{\rm ff}$ at constant $\sigma_f(^{238}$U) = $0.014\, \mathrm{b}$.  We compute 368 networks in (a) and (b) and 483 networks in (c); the jagged regions are an artifact of the finite resolution of the grid. Scenarios A, B, C, and D from Table~\ref{Table3} are indicated in the panels.  The Red region approximately reaches carbon ignition \cite{1992ApJ...396..649T}.}	
\end{figure*} 

\begin{figure*}[ht!]
\centering  
\includegraphics[trim=0 105 310 45,clip,height=5cm,keepaspectratio]{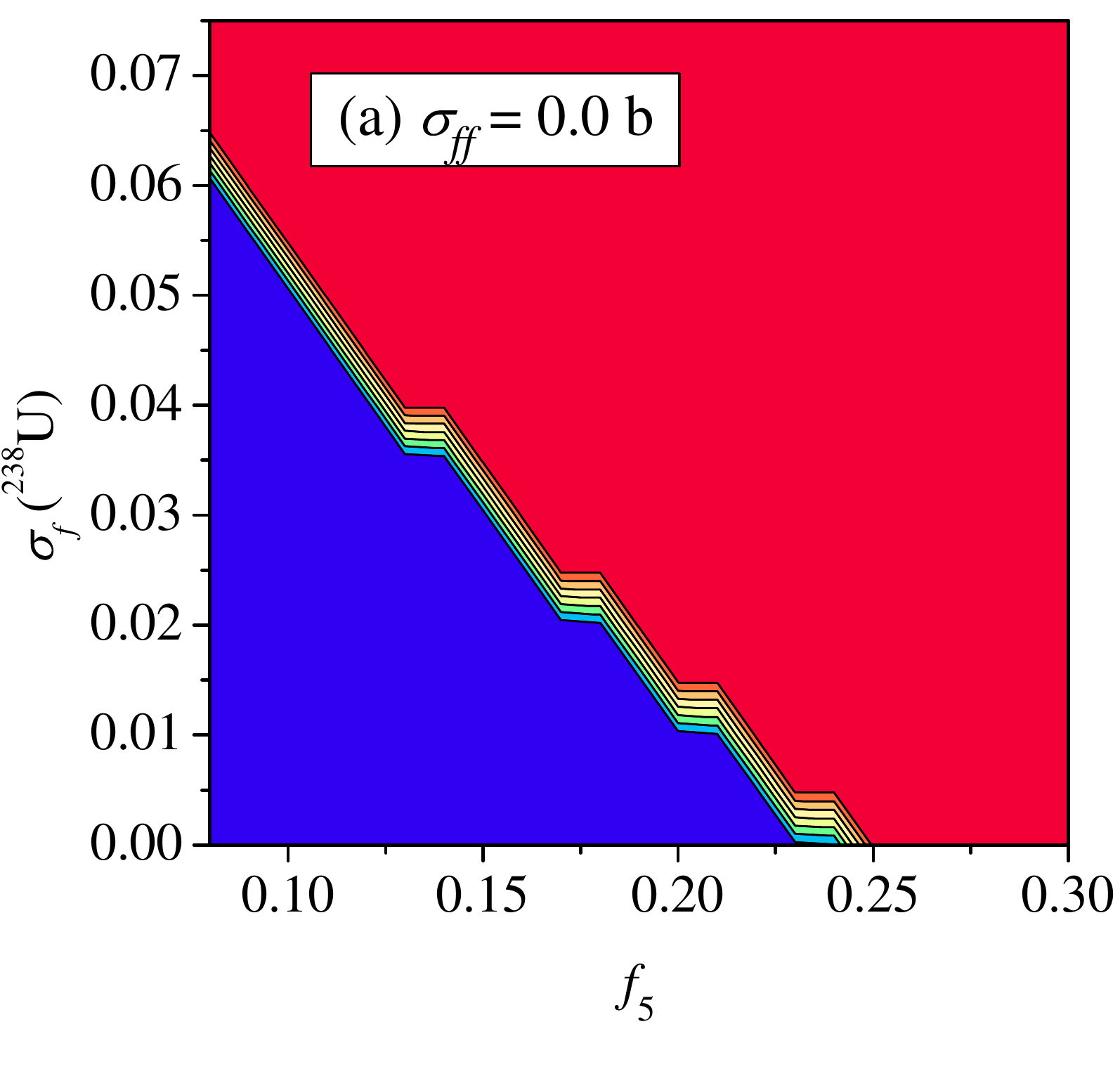} 
\includegraphics[trim=0 105 310 45,clip,height=5cm,keepaspectratio]{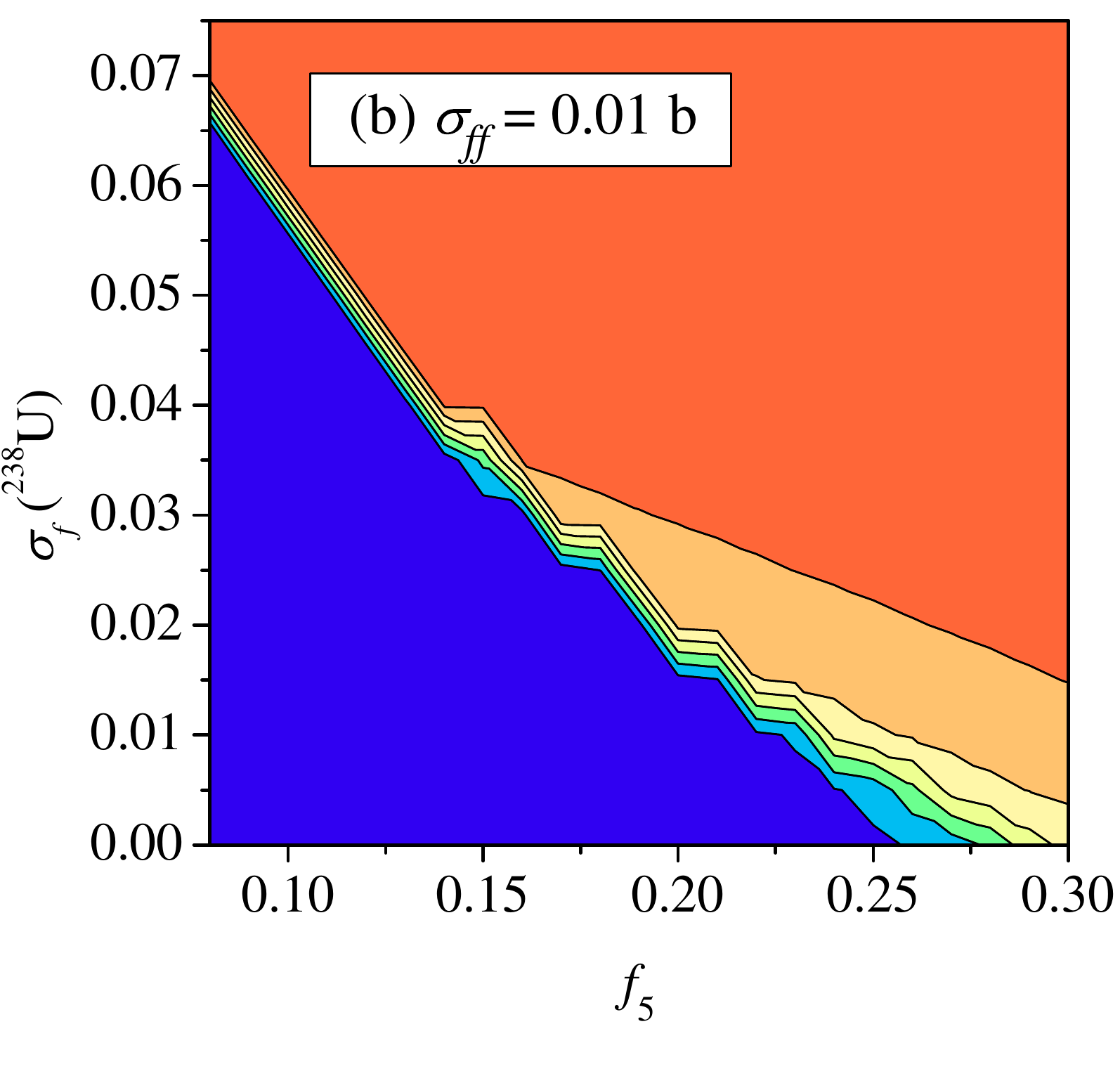}
\includegraphics[trim=0 105 213 45,clip,height=5cm,keepaspectratio]{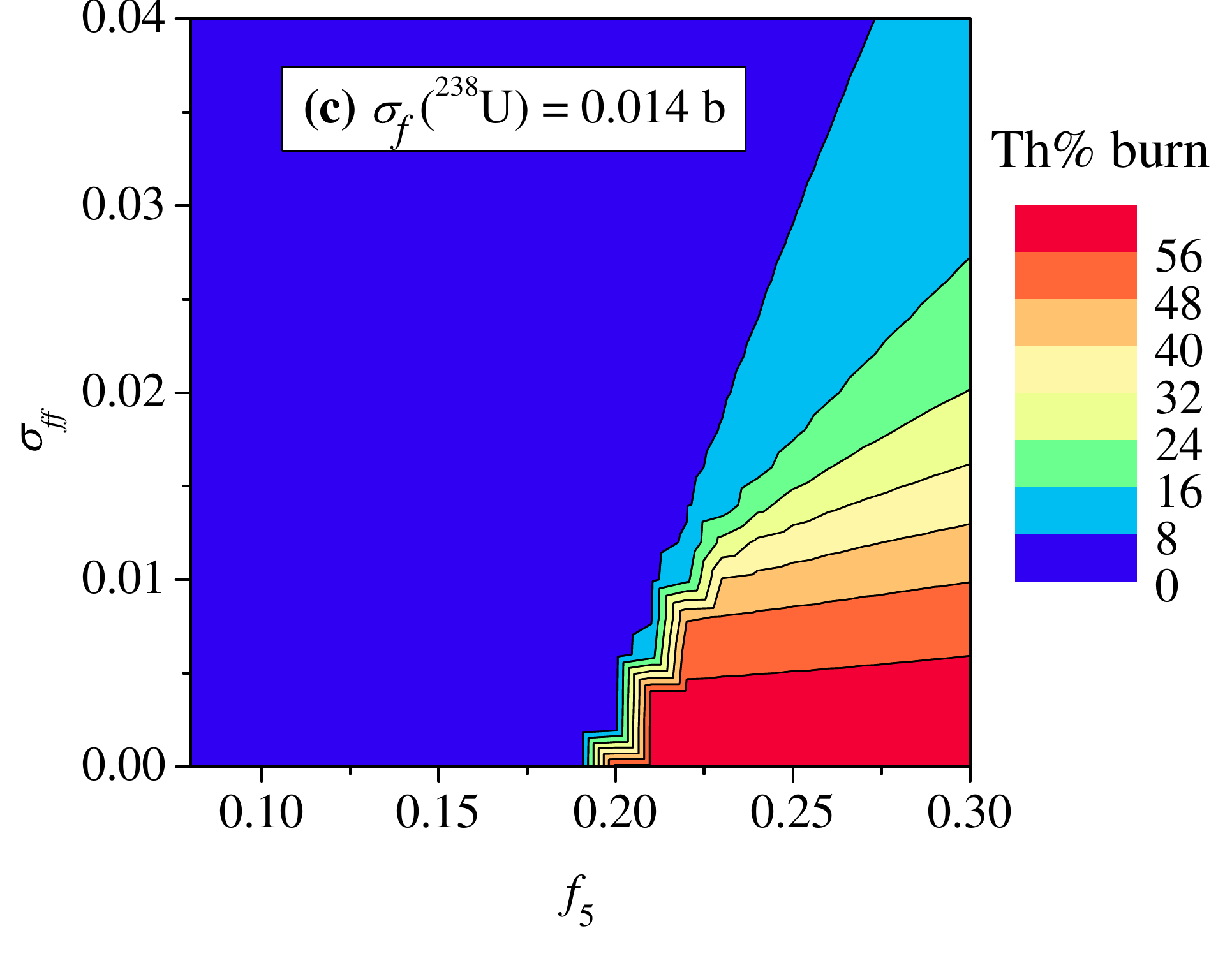}
\includegraphics[trim=0 105 310 45,clip,height=5cm,keepaspectratio]{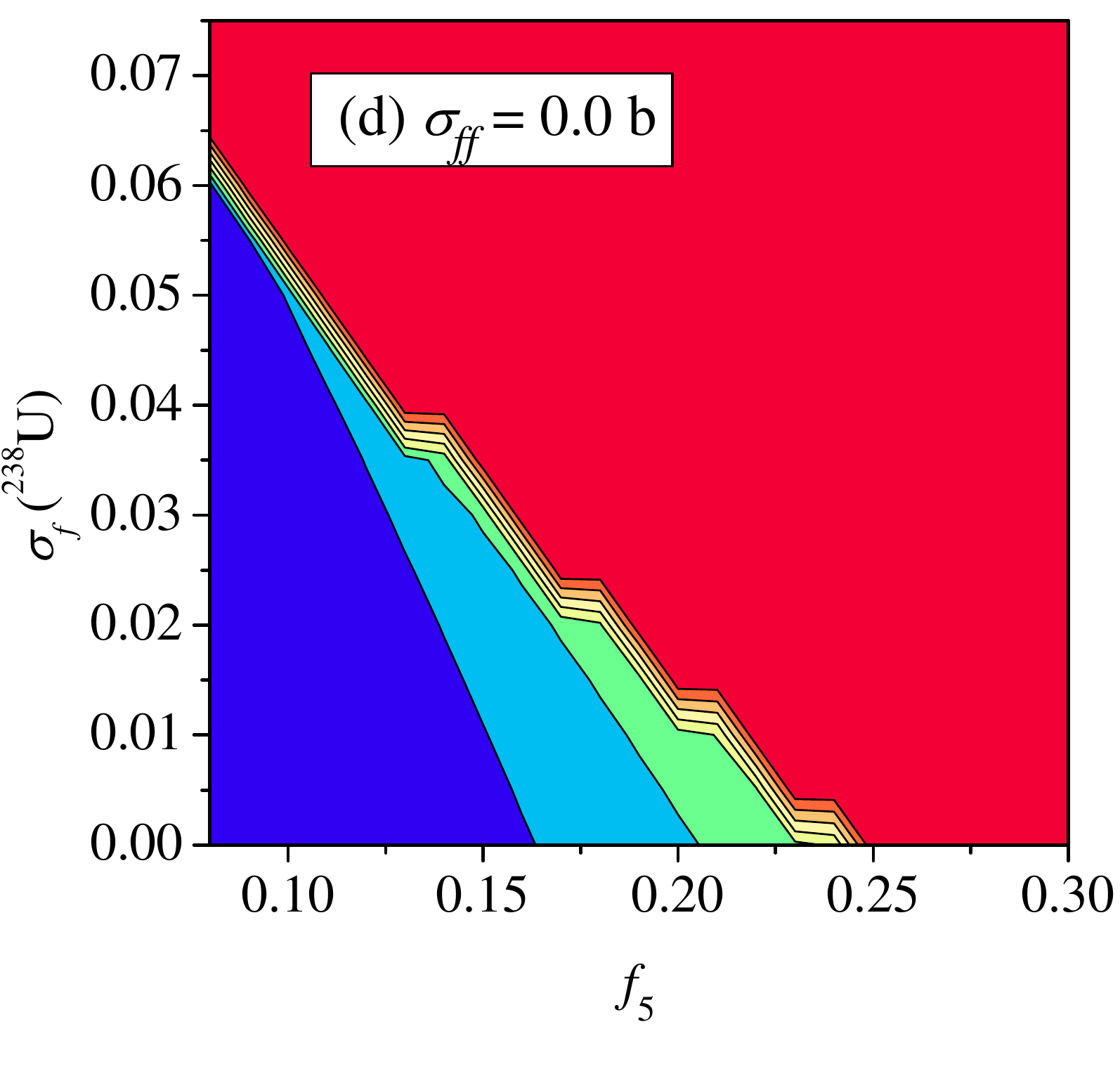} 
\includegraphics[trim=0 105 310 45,clip,height=5cm,keepaspectratio]{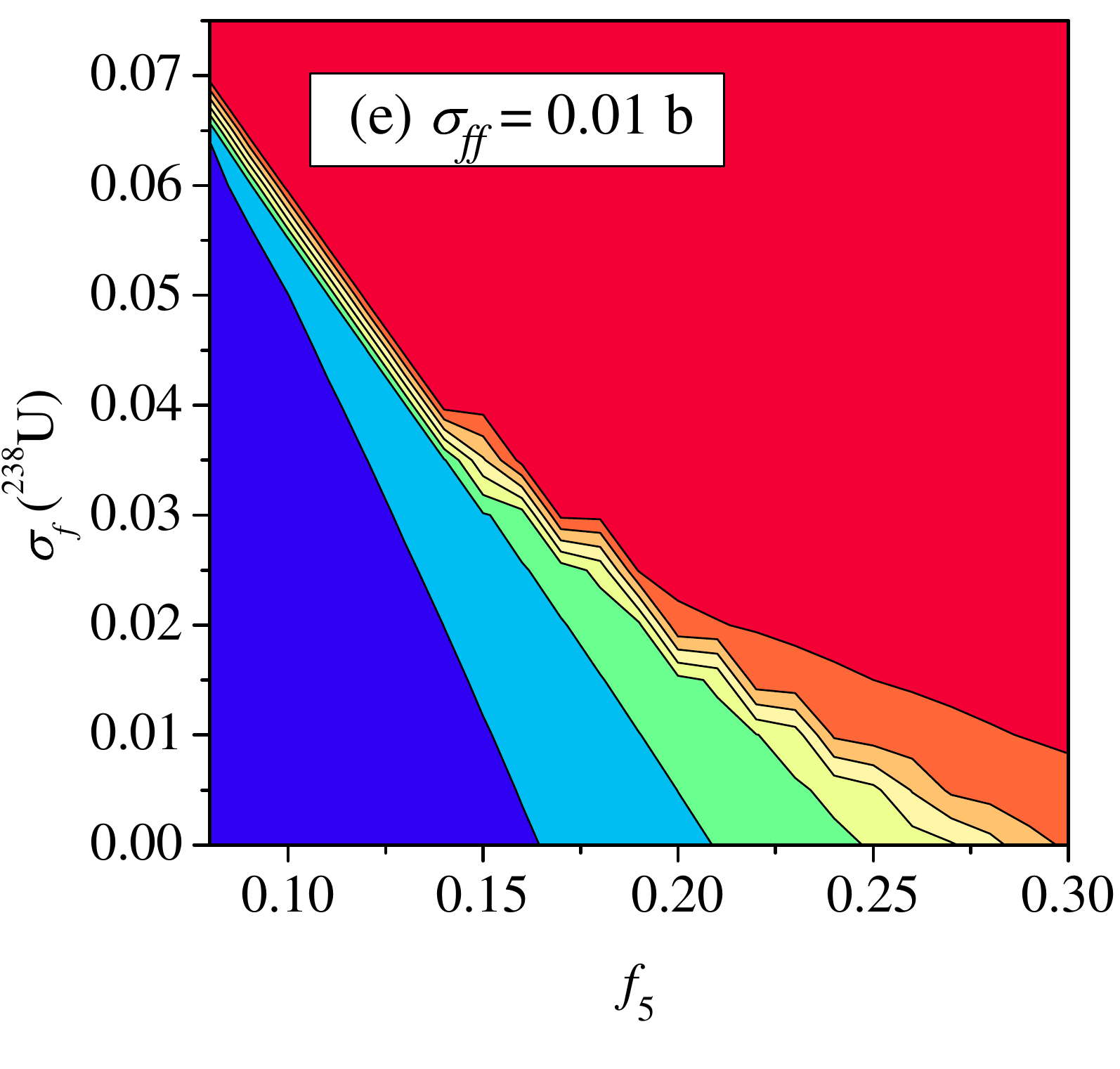}
\includegraphics[trim=0 105 213 45,clip,height=5cm,keepaspectratio]{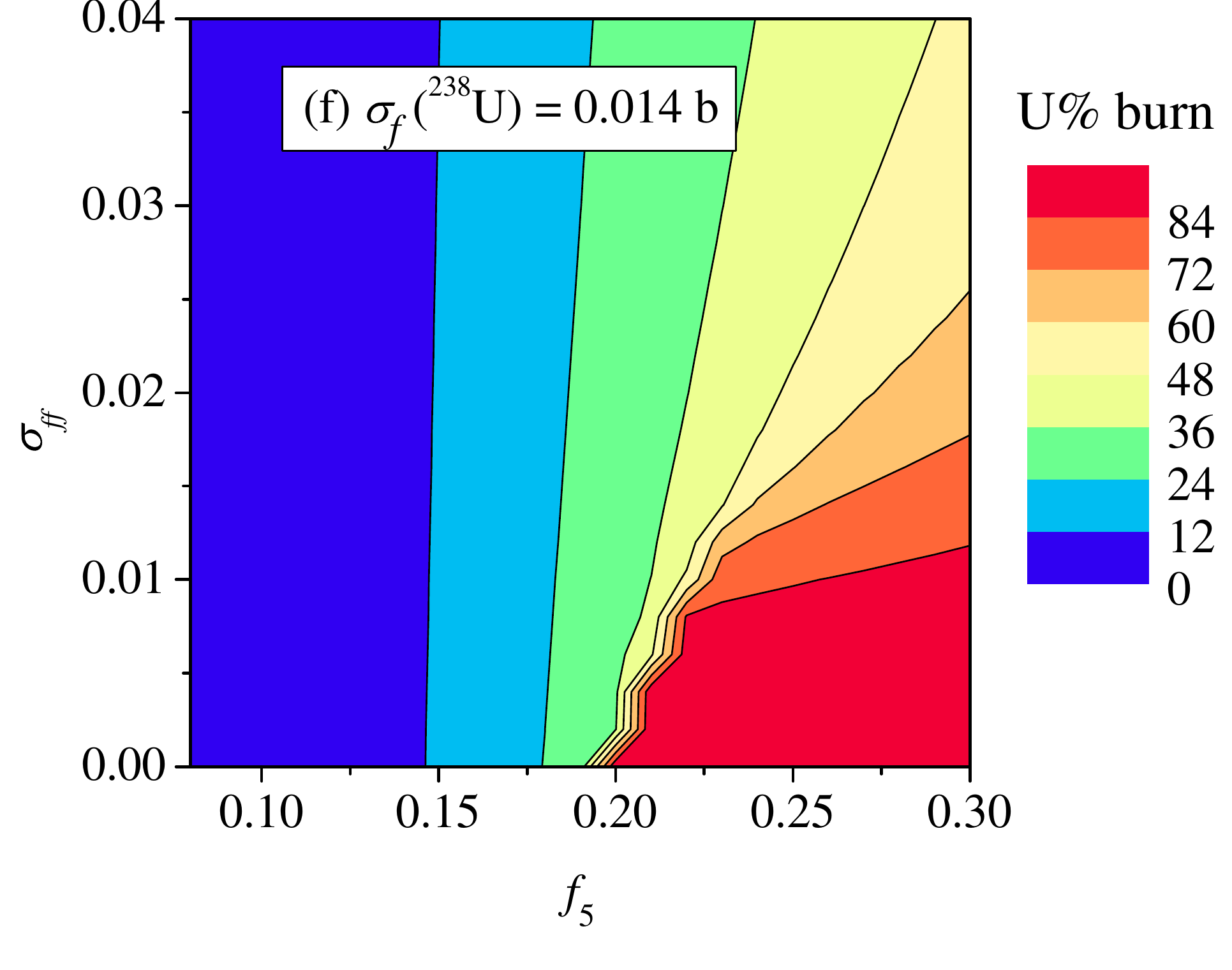}
\caption{\label{Fig7} Percentage of Th that fissions (top) and total fraction of U that fissions (bottom) for the same simulations as in Fig. \ref{Fig6}.}	
\end{figure*} 

\subsection{Heat release and final temperature}

The heating rate for the different scenarios is shown in Fig. \ref{Fig2} and the total heating is plotted in Fig. \ref{Fig3} and listed in Table \ref{Table3}. The final temperature is plotted in Fig. \ref{Fig6} for a range of $f_5$ values.  There is a minimum value of $f_5$ for the system to be critical.  Below this value there is almost no fission heating and $T_f$ is small.  Next there is a modest range of $f_5$ values where significant $^{235}$U fissions but little $^{238}$U or Th fissions.  Here $T_f$ is between $\approx 1\textrm{--}2\times 10^9\, \mathrm{K}$.  Finally there is a sharp transition when there are enough neutrons for an essentially complete breeder reaction stage that fissions most of the U and 58\% of the Th. 

Note that the breeder reaction stage leads to a large total energy release of $\approx 0.36\, \mathrm{MeV} / \mathrm{nucleon}$ as listed in Table~\ref{Table3}; this leads to $T_f\approx 6\times 10^9\, \mathrm{K}$.  As the $^{238}$U fission cross section increases, the necessary $f_5$ for the breeder reaction decreases. In general, the breeder reaction uses all available fuel resulting in a nearly complete burn. 

According to Fig. 6 of Timmes and Woosley \cite{1992ApJ...396..649T} a final temperature of $T_f \approx 6\times 10^9\, \mathrm{K}$ for a $5\, \mathrm{mg}$ mass may be hot enough to ignite carbon burning.  However, hydrodynamical simulations should be performed to explicitly verify that the energy release $S$ heats the system enough to start carbon burning, which we reserve for future work. We note that the presence of Pb in Table \ref{Table1} significantly increases the heat capacity without increasing the fission energy released.  If the amount of Pb were less (or absent) the system would reach higher temperatures.

\subsection{Cross section and enrichment sensitivity}

To explore sensitivity to input parameters we have performed large numbers of reaction network simulations.  The frames in Fig. \ref{Fig6} were prepared by computing a grid of networks to find the final temperature as a function of $f_5$, $\sigma_f(^{238}\mathrm{U})$, and $\sigma_{\rm ff}$. 
In Fig. \ref{Fig6}a and \ref{Fig6}b we compare the final temperature when excluding neutron-captures on the fission fragments (i.e., $\sigma_{\rm ff} = 0.0 \, \mathrm{b}$) and assuming a modest $\sigma_{\rm ff} = 0.01\, \mathrm{b}$ as an average neutron-capture cross section, respectively. The grid is computed between $f_5 =0.08$ and $f_5 =0.30$ at a resolution of $\Delta f_5 = 0.01$, and between $\sigma_f(^{238}\mathrm{U}) =0.00 \, \mathrm{b}$ and $0.075\, \mathrm{b}$ with a resolution of $\Delta \sigma_f(^{238}\mathrm{U}) = 0.005\, \mathrm{b}$, for a total of 368 network calculations. The  cases A, B, C, and D from Table~\ref{Table3} are marked in Fig.~\ref{Fig6}. 

In Fig. \ref{Fig6}a, where no neutron-captures occur on the fission fragments, there is a sharp transition between the incomplete burn or `fizzle' behavior at low $f_5$ and the complete burn at high $f_5$ where the final temperature is $T_f \gtrsim 5 \times 10^9 \, \mathrm{K}$. For enrichments $f_5 \lesssim 0.25$ the transition between incomplete/complete burns depends on $\sigma_f(^{238}\mathrm{U})$, requiring larger $\sigma_f(^{238}\mathrm{U})$ for a complete burn at lower enrichment $f_5$. For enrichments $f_5 \gtrsim 0.25$ a nonzero $\sigma_f(^{238}\mathrm{U})$ always results in a complete burn. In Fig. \ref{Fig6}b the fission fragments act as a neutron sink with a cross section of $\sigma_{\rm ff} = 0.01\, \mathrm{b}$.  In this case, the burning transition shifts to higher $f_5$. However, at nonzero $\sigma_f(^{238}\mathrm{U})$ we still find a robust burn of the $^{238}\mathrm{U}$ and Th. 

In Fig. \ref{Fig6}c we explore the sensitivity to the fission fragments as a neutron sink assuming constant $\sigma_f(^{238}\mathrm{U})$=0.014 b. We use a resolution of $\Delta \sigma_{\rm ff} = 0.002$ b; the resolution in $f_5$ is the same as above, for a total of 483 networks.
For low values of $\sigma_{\rm ff}$ (\ie\ $\sigma_{\rm ff} \lesssim \sigma_f(^{238}\mathrm{U})$), we find that complete burns are still readily achieved for $f_5 \gtrsim 0.20$. It is only at $\sigma_{\rm ff} \gtrsim \sigma_f(^{238}\mathrm{U})$ that the fission fragments begin to `outcompete' the $^{238}\mathrm{U}$ for neutrons and quench the burning. Thus, even with some degree of `poisoning' due to neutron captures onto fission fragments, or other impurities, the system may still undergo a complete burn so long as $\sigma_f(^{238}\mathrm{U})$ (and $f_5$) is sufficiently high.

In Fig. \ref{Fig7} we show the total percentage of Th and U that fissions in the grid of networks computed in Fig. \ref{Fig6}. It is clear that barely any Th (top) burns in networks that fizzle at low $f_5$ (Fig. \ref{Fig7}a), and that the conditions required for a complete burn Th burning have a sharp transition. Neutron captures on the fission fragments can suppress the total Th fraction that fission (Figs. \ref{Fig7}b and \ref{Fig7}c), and the sharp turn-on is shifted to slightly greater $f_5$. When we compare to the U fraction that burns (bottom) in the networks that fizzle, we see that most of the $^{235}$U burns, but barely more than $f_5$, so the heating is largely due to a $^{235}$U burning which then stalls before igniting the breeder stage. We conclude that the final temperatures observed in Fig. \ref{Fig6} can be explained by a steady increase in $^{235}$U burning with increasing $f_5$ in the fizzling regime, with complete burns achieved after a very sharp turn-on which burns the Th and $^{238}$U in a breeder stage.



\section{Discussion}
In this section we discuss limitations in our reaction network and then carbon ignition.

\subsection{Limitations of reaction network}

We now explore possible limitations in our reaction network.  First we have only included $(n,\gamma)$ and fission reactions.  These reactions proceed very rapidly on a timescale of $\tau \sim 10^{-15} \, \mathrm{s}$ and neglecting beta-decay and $(\gamma,n)$ should be a good first approximation.  We have neglected $(n,2n)$ reactions because these should be unimportant except at high neutron energies above $1\, \mathrm{MeV}$. In future work we will examine temperature-dependent cross sections and any new reaction pathways that may result.

Our reaction network assumes $1 \, \mathrm{MeV}$ neutrons.  This is a reasonable first approximation to the fission spectrum as long as the amount of light nuclei such as C or O is small.  If light nuclei are present, then nuclear recoil following elastic scattering will reduce the neutron energies.  In future work we will explore sensitivity of the reaction network to the neutron spectrum.


Our reaction network is somewhat incomplete and does not include reactions for very neutron-rich Th or U isotopes.  For U we include reactions on isotopes up to $^{241}$U.  The omission of reactions on $^{242}$U or heavier isotopes is not expected to be important because most of the U fissions and only a tiny fraction captures enough neutrons to reach $^{242}$U.  

For Th we only include reactions on $^{232}$Th and $^{233}$Th.  For Cases B, C and D a significant fraction of the original Th is converted to $^{234}$Th (which is stable in our network).  Including reactions on heavier Th isotopes could lead to the fission of more Th.  In particular, the neutron-induced fission of $^{235}$Th could be a significant addition to our reaction network. There may not be data for this neutron-rich isotope, however, and further progress using our reaction network likely will require theoretical rates for very neutron-rich isotopes.

We assumed a constant baryon density in Sec.~\ref{sec.formalism} Eqs. \ref{eq.dydt},\ref{eq.dyffdt}, and \ref{eq.dyAdt}.  In reality, the system will expand slightly because of the large fission energy release.  However this decrease in density is only about 25\% because the electrons are degenerate \cite{fission2}.  This will slightly slow down the rate of all neutron reactions and therefore the chain reaction will take slightly longer to complete. 

We have neglected cooling from heat conduction and neutrino emission.  This should be a good approximation during the fission reaction because the reaction rates are so high.  The fission heating rate in Fig. \ref{Fig2} is consistent with the estimated rate in Ref. \cite{fission2}.  The rate of cooling from heat conduction via the large thermal conductivity of the degenerate electrons is estimated in Ref. \cite{fission2} to be two to three orders of magnitude lower than the heating rate in Fig.~\ref{Fig2}.  Therefore heat conduction is unimportant during the fission reaction.

\subsection{Carbon ignition and explosive yield}

We now consider carbon ignition.  After the fission chain reaction, the system may be so hot that self-propagating thermonuclear carbon burning is initiated. Such a scenario is unstudied and future hydrodynamical simulations will be needed. In the meantime, however, we can examine the fission-scenario through analogy with terrestrial nuclear weapons.

First for context, we discuss ignition of hydrogen isotopes in a terrestrial nuclear weapon.  The Classical Super was the original idea to use heat from an atomic bomb to start fusion in a deuterium, tritium mixture.  It is said that losses from thermal radiation will help quench thermonuclear fusion in the Classical Super.  Instead, radiation implosion can be used to compress the hydrogen fuel first.  This compression increases the energy density from hydrogen fusion without increasing the radiation losses.  In a WD, the carbon fuel is already at a very high density.  Therefore radiation losses are likely unimportant and our system may avoid this problem with the Classical Super.  Because of the high initial density there may be no need for radiation implosion to compress the system further. 

It is interesting to compare the explosive yields in our system to those of nuclear weapons.  Previously we had estimated the initial mass of the uranium rich crystal to be about $5 \, \mathrm{mg}$ \cite{PhysRevLett.126.131101,fission2}.  In Cases B, C, and D we find a significant fraction of the U and Th in this 5 mg mass fissions.  This will release energy equivalent to about $50\, \mathrm{kg}$ of TNT.  The yield is much lower than the approximately $15 \, \mathrm{kiloton}$ yield of the first atomic bombs because the $5\, \mathrm{mg}$ mass of our system is much less than the multi kg core masses of conventional fission weapons.  Note that the efficiency of our system may be much higher, with nearly all of the U and Th fissioning, compared to the few \% efficiency of an atomic bomb.  Nevertheless, because the mass and critical mass are so much smaller, our fission yield is almost a million times smaller.  Therefore, if carbon burning is not initiated, the fission chain reaction may have very little effect on the star.  

The situation is dramatically different if carbon burning is initiated.  The thermonuclear burning of a significant fraction of C in a WD will release energy comparable to a SN Ia.  This corresponds to an explosive yield of almost $10^{29}\, \mathrm{megatons}$ (MT)!  Thus, we propose using a $50\, \mathrm{kg}$ yield fission {\it primary} to ignite a  $10^{29} \, \mathrm{MT}$ fusion {\it secondary}.

The temperature required for carbon ignition depends on the ignition scenario including the system size, density and ignition timescale.  Our system has a density near $10^8 \, \mathrm{g \ cm^{-3}}$ and a total mass of order $5 \, \mathrm{mg}$.  The original work of Timmes and Woosley may be the most directly relevant previous calculation of ignition temperature for our conditions \cite{1992ApJ...396..649T}.  They consider ignition in a C/O liquid where they instantaneously replace a small mass of C/O by carbon burning ashes and assume the temperature of this ash has been raised to $T_i$.  They find an ignition temperature of $T_i\approx 5\times 10^9\, \mathrm{K}$.  If the final temperature $T_f>T_i$ a flame may propagate in the surrounding C/O liquid.  If $T_f<T_i$ the system will cool via heat conduction without initiating carbon burning.  This ignition temperature of $5\times 10^9\, \mathrm{K}$ should be verified with future hydrodynamic simulations.





\section{Conclusions\label{sec.conclusions}}

In this paper we have performed novel reaction network simulations of fission chain-reactions in a cooling WD.  The first solids to form when material in a WD just starts to crystallize are expected to be U and Th rich because of their high charges.  These solids may support a fission chain-reaction if the uranium enrichment $f_5$ is high enough $> 0.12$.

We find that the reaction proceeds very quickly (within $\approx 10^{-11}$ s) because the density is high and the neutron cross sections are large. In general, the reaction proceeds in two stages.  In the first stage neutrons from $^{235}$U fission transform or breed some $^{238}$U and $^{232}$Th nuclei into more easily fissionable $^{239}$U and $^{233}$Th.  In the second, or breeder reaction, stage most of the original U and a significant fraction of the Th fission.  These reaction stages release $\approx 0.36 \, \mathrm{MeV / nucleon}$ and raise the final temperature to $T_f \approx 6 \times 10^9 \, \mathrm{K}$. This is important input for our thermal diffusion simulations where we find that carbon ignition is likely at high densities \cite{bombcode}.




{\it Acknowledgements:} 
We thank Ezra Booker, Constantine Deliyannis, Erika Holmbeck, Wendell Misch, Matthew Mumpower, Witek Nazarewicz, Catherine Pilachowski, Tomasz Plewa, and Rebecca Surman for helpful discussions.  The work of CJH was performed in part at the Aspen Center for Physics, which is supported by National Science Foundation grant PHY-1607611. This research was supported in part by the US Department of Energy Office of Science Office of Nuclear Physics grants DE-FG02-87ER40365 and DE-SC0018083 (NUCLEI SCIDAC).

\providecommand{\noopsort}[1]{}\providecommand{\singleletter}[1]{#1}%
%



%



%
\end{document}